\documentclass{aastex}
\usepackage{emulateapj5,times,mathptm}
\singlespace
\slugcomment{Accepted for the publication in the {\it Astrophysical Journal Letters}}
\def\farcm{\hbox{$.\mkern-4mu^\prime$}}
\def\farcs{\hbox{$.\mkern-4mu^{\prime\prime}$}}

\def\la{\mathrel{\hbox{\rlap{\hbox{\lower4pt\hbox{$\sim$}}}\hbox{$<$}}}}
\def\ga{\mathrel{\hbox{\rlap{\hbox{\lower4pt\hbox{$\sim$}}}\hbox{$>$}}}}

\shortauthors{Park}
\shorttitle{N49B}

\begin{document}
\title{Detection of Magnesium-Rich Ejecta in the Middle-Aged Supernova 
Remnant N49B}

\author{Sangwook Park\altaffilmark{1},
John P. Hughes\altaffilmark{2}, Patrick O. Slane\altaffilmark{3},
David N. Burrows\altaffilmark{1}, Jessica S. Warren\altaffilmark{2},
Gordon P. Garmire\altaffilmark{1}, and John A. Nousek\altaffilmark{1} }

\altaffiltext{1}{Department of Astronomy and Astrophysics, Pennsylvania State
University, 525 Davey Laboratory, University Park, PA. 16802; park@astro.psu.edu}
\altaffiltext{2}{Department of Physics and Astronomy, Rutgers University,
136 Frelinghuysen Road, Piscataway, NJ. 08854-8019}
\altaffiltext{3}{Harvard-Smithsonian Center for Astrophysics, 60 Garden Street,
Cambridge, MA. 02138}

\begin{abstract}

The middle-aged supernova remnant (SNR) N49B in the Large Magellanic 
Cloud has been observed with the {\it Chandra X-Ray Observatory}. The 
superb angular resolution of {\it Chandra} resolves the complex structure 
of X-ray emitting filaments across the SNR. All observed features are 
soft ($E <$ 3 keV) and we find no evidence for either point-like or 
extended hard emission within the SNR. Spectral lines from O, Ne, Mg, Si, 
S, and Fe are present. Equivalent width images for the detected elemental 
species and spatially-resolved spectral analysis reveal the presence of 
Mg-rich ejecta within the SNR. We find no such enrichment in O or Ne, 
which may reflect details of the nucleosynthesis process or the heating 
and cooling of the ejecta as it evolved. The bright circumferential 
filaments are emission from the shocked dense interstellar medium (ISM). 
We detect faint diffuse X-ray emission that extends beyond the X-ray 
bright filaments toward the west and southeast. These features appear to 
be the blast wave shock front expanding into lower density portions of the 
ISM seen in projection. We set an upper limit of $\sim$$2\times 10^{33}$ 
ergs s$^{-1}$ on the 0.5 $-$ 5 keV band X-ray luminosity of any embedded 
compact object.

\end{abstract}


\section {\label {sec:intro} INTRODUCTION}

N49B (0525$-$66.0) is a supernova remnant (SNR) in the Large
Magellanic Cloud (LMC), which is located $\sim$6$\farcm$5 northwest of
N49, another bright LMC SNR \citep{milne72}. In the optical band, N49B
consists of two irregular-shaped knots, separated by $\sim$1$\farcm$4 
\citep{mathewson83}; spatially variable extinction is present around
this SNR. The {\it Einstein}/HRI and {\it ROSAT}/HRI data revealed a 
nearly circular X-ray morphology encompassing the optical knots 
\citep{mathewson83,williams99}. Radio data also showed a circular 
morphology and a steep radio spectral index ($\alpha$ $\sim$ $-$0.7 
where $S_\nu\propto \nu^\alpha$), typical for shell-type SNRs 
\citep{dickel98}. Chu \& Kennicutt (1988) suggested a Population I 
progenitor for N49B based on its proximity to the H{\small II} region 
DEM 181 and a possible association with a molecular cloud \citep{cohen88}.

The {\it ASCA} spectrum of N49B was described with a nonequilibrium
ionization (NEI) Sedov model with {\it kT} $\sim$ 0.4 keV \citep{hughes98}. 
The fitted metal abundances were consistent with the LMC interstellar 
medium (ISM), although there was evidence for a slightly enhanced Mg 
abundance. The measured X-ray absorbing column ($\sim$2.5 $\times$ 
10$^{21}$ cm$^{-2}$) was in agreement with the H{\small I} column toward 
N49B ($\sim$$2 \times 10^{21}$ cm$^{-2}$) \citep{heiles79,mcgee66}. This 
relatively high column, as compared with other LMC SNRs, supports N49B's 
physical association with a moderately dense ISM ($n_0$ $\sim$ 0.8 
cm$^{-3}$ and swept-up mass of $\sim$560 $M_{\odot}$; Hughes et al. [1998]). 
A discrepancy between the ionization and dynamical ages ($\sim$3 kyr vs. 
$\sim$10 kyr) was attributed to the SN explosion having occurred in a 
pre-existing interstellar cavity \citep{hughes98}. {\it Einstein} and 
{\it ROSAT} images revealed the broadband X-ray structure of N49B down 
to $\sim$5$^{\prime\prime}$ angular scales. {\it ASCA} data provided an 
integrated spectrum, but to date spatially-resolved spectroscopy has been 
impossible because of the poor angular resolution of {\it ASCA} and the 
small angular size of N49B. The superb angular resolution of the {\it 
Chandra X-Ray Observatory} allows us to resolve the complex structure of 
N49B with good photon statistics and spectral resolution. Here we report 
results from high resolution image and spatially-resolved spectral analyses 
of the {\it Chandra} observation of N49B.

\section{\label{sec:obs} OBSERVATION \& DATA REDUCTION}

N49B was observed with the Advanced CCD Imaging Spectrometer (ACIS) on
board {\it Chandra} on 2001 September 15 as part of the Guaranteed Time
Observation program. The ACIS-S3 chip was chosen to utilize the best
sensitivity and energy resolution of the detector in the soft X-ray
band. The pointing was toward the X-ray centroid of N49B
($\alpha$$_{2000}$ = 05$^{h}$ 25$^{m}$ 26$^{s}$.04, $\delta$$_{2000}$
= $-$65$^{\circ}$ 59$'$ 06$\farcs$9) with the aim point shifted by 245
detector pixels ($\sim$2$'$) from the nominal aim point on S3 in order
to simultaneously observe the nearby LMC SNR N49 on the same CCD chip
as well.  Results from the analysis of the N49 data are presented
elsewhere \citep{park03}.

We utilized the data reduction techniques developed by Townsley
et al.~(2002a) for correcting the spatial and spectral degradation of
the ACIS data caused by radiation damage, the charge transfer 
inefficiency (CTI; Townsley et al.~2000). We screened the data with 
the flight timeline filter and then applied the CTI correction before 
further standard screening by status and grade. There is no 
significant photon flux from N49B at $E$ $\ga$ 3 keV (above background), 
thus we only use X-rays with $E$ $<$ 3 keV in the subsequent analysis. 
The overall lightcurve was examined for possible contamination from 
time-variable background. The background flux is relatively high
(by a factor of 2) for the last 10\% of the exposure. We have excluded 
the 4 ks time interval affected by this flare. This data reduction
results in effective exposure of 30.2 ks which yields 143,000 photons 
in the 0.3$-$3 keV band from the SNR within a 1$\farcm$4 radius region.

\section{\label{sec:image} X-RAY IMAGES}

\begin{figure}[h]
\figurenum{1}
\centerline{\includegraphics[angle=0,width=3.5in]{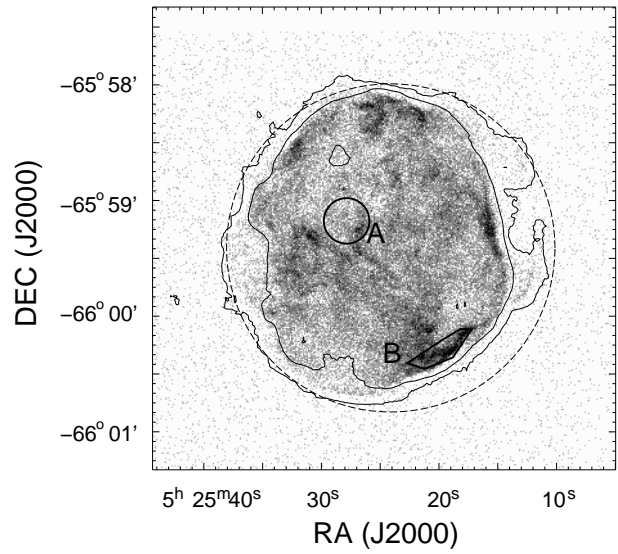}}
\figcaption[]{ The broadband (0.3 $-$ 3.0 keV) ACIS image of N49B. Dark 
gray-scales indicate higher intensities. The regions we use for the 
spectral analysis are marked as ``A'' and ``B''. In order to show the
faint diffuse emission that extends beyond the bright filaments, contours 
of the low X-ray surface brightness from the smoothed broadband image are 
overlaid. The dashed circle (1$\farcm$4 radius) displays a fiducial boundary 
for the SNR, which is also roughly coincident with the radio boundary.
\label{fig:fig1}}
\end{figure}

Figure \ref{fig:fig1} is the broadband ACIS image of N49B. The ACIS
image reveals X-ray filamentary structures down to subarcsec angular 
scales. The outermost and faintest X-ray emission traces a roughly 
circular boundary ($\sim$1$\farcm$4 radius). The brightest parts of the 
circumferential filaments (notably along the southwest, west, north, and 
northeast) appear to be flattened. The surface brightness of the complex 
filamentary structures in the middle of the eastern side is nearly as 
high as at the bright rims. Elsewhere in the interior the emission 
is more than an order of magnitude fainter than at the rim. The faint 
radio shell that extends toward the west \citep{dickel98}, which
has not been seen in previous X-ray observations, is detected beyond the 
bright X-ray filament in the west (Figure \ref{fig:fig1}). This X-ray 
``double-shell'' structure in the west, as well as the faint outermost 
``plateau'' of emission toward the southeast, may be projection effects 
of the shock propagating into low density regions of the ISM, with the 
bright filaments corresponding to regions where dense ISM has been 
encountered (more discussion in \S \ref{sec:disc}).

The surface brightness distribution of N49B is dominated by X-ray emission 
below 3 keV and we find no evidence for an embedded hard point source or 
diffuse synchrotron nebula. We do detect spectral variations across the SNR: 
emission from the northeast is softer with a larger relative flux at 
$E$~$<$~0.8 keV, while emission in the central-to-west regions is harder 
(enhanced at 1.1~$<~E~<$~3.0 keV). These variations in the broadband X-ray 
fluxes appear to imply variations in the absorbing column and/or local 
enhancements of line emission, especially from Mg and Si. To 
quantify the column density variations, we compare the spectra of numerous 
small angular regions to the SNR's total spectrum, following the method 
described in Warren et al. (2003). The average column density measured by 
{\it ASCA} over the entire SNR is $N_H$ = 2.5 $\times$ 10$^{21}$ cm$^{-2}$ 
\citep{hughes98}. Spectra from the bright northern limb indicate column 
densities lower than the average by $\sim$$1.4\times 10^{21}$ cm$^{-2}$. 
Spectra from the southwest suggest columns higher than the average by up 
to $\sim$$2\times 10^{21}$ cm$^{-2}$. This means a factor of $\sim$4 
variation of the absorbing column across N49B.

We explore the angular distributions of the line emission by constructing 
{\it equivalent width} (EW) images for the detected elemental species, 
following the method by Park et al. (2002). EW images for O, Ne, and Mg 
were generated by selecting photons around the broad line features. The 
O and Ne EW images are featureless, which indicates the {\it absence} of 
any local enhancements of O/Ne line emission. The Mg EW image, on the 
other hand, reveals significant enhancements near the center of the SNR 
(Figure \ref{fig:fig2}), establishing that the hard enhancements in the 
central-to-western regions of the SNR are caused by enhanced Mg line 
emission. The Mg EW distribution suggests substantial variations in the 
metal abundances and/or temperature/ionization between the center and 
periphery of the SNR. The bright circumferential X-ray filaments are 
featureless in all EW maps indicating that there is little azimuthal 
variation of abundance and/or thermodynamic state around the rim.

\section{\label{sec:spec} X-ray Spectra} 

In order to investigate the origin of the line enhancements, we
extracted spectra from a small region where the Mg EW is high (region
A in Figures \ref{fig:fig1} and \ref{fig:fig2}), and a bright X-ray 
filament (region B; Figure \ref{fig:fig1}). The extracted source spectra 
(Figure \ref{fig:fig3}) contain $\sim$3700 (region A) and $\sim$7300 
(region B) photons in the 0.4 $-$ 3.0 keV band. Both spectra were binned 
to contain a minimum of 20 counts per channel. For the spectral analysis 
of our CTI corrected data, we used the response matrices appropriate for 
the spectral redistribution of the CCD \citep{townsley02b}. The low 
energy ($E$~$<$~1 keV) quantum efficiency (QE) of the ACIS has degraded 
with time due to accumulating molecular contamination on the optical 
blocking filter. We have corrected the time-dependent QE degradation by 
modifying the ancillary response function for the extracted spectrum 
using the {\it ACISABS} software \citep{chartas03}.

We fit the spectra with an NEI plane-parallel shock model 
\citep{borkowski01}. Elemental abundances were fixed for He (0.89), 
C (0.30), N (0.12), Ca (0.34) and Ni (0.62) at the LMC values 
\citep{russell92} (hereafter, all abundances are with respect to solar; 
Anders \& Grevesse [1989]) since our data do not constrain the
contribution from these species. Other elemental abundances were
allowed to vary freely. The region A spectrum reveals a remarkably
strong Mg line with a best-fit abundance of 1.7. This is more than 5 
times higher than the LMC abundance. Si is also overabundant (1.4 or 
$\sim$5 times the LMC abundance). O, Ne, and Fe are best fitted with 
LMC-like abundances. In contrast to region A, the region B spectrum is 
best fitted with the LMC abundances for O, Ne, and Mg (Table 
\ref{tbl:tab1}). Other bright circumferential filaments show 
similar spectral features and derived abundances to region B.These 
indicate that the X-ray bright filaments are dominated by emission from 
the shocked ISM. We thus confirm that the EW distributions are primarily 
caused by abundance variations. 

The use of a single plane shock model to describe region A is 
oversimplistic, since this position contains emission from
shock-heated swept-up ISM integrated through the depth of the SNR, 
besides a possible ejecta component. Although the single-temperature 
plane shock model indicates a high Mg abundance as the primary cause 
of the strong Mg line, we also test whether higher temperature plasma 
in the SNR interior could be the source of the strong Mg line. For 
this purpose, we use a Sedov model, which provides a physically motivated 
temperature distribution, to fit the region A spectrum. Even these fits, 
however, require a high Mg abundance ($\sim$1.4) and LMC-like abundances 
for O and Ne ($\sim$0.2). We finally employ a two-component model 
consisting of a Sedov component to account for the shocked ISM and a 
plane shock component to describe the metal-rich ejecta. We set the 
Sedov model parameters to those previously obtained by {\it ASCA} 
\citep{hughes98}, since they are consistent with our own Sedov fit 
results, and fix abundances at LMC values. Fits with no constraint on 
the model normalizations result in a negligible contribution from the 
Sedov model, i.e., the single-temperature plane shock model (with 
enhanced Mg) is the preferred model. Even if we force each component to 
contribute equally to the observed emission, the high Mg abundance 
(with low O and Ne) persist in the ejecta component. In all cases, 
the best-fit temperatures are $kT \sim 0.4$ keV. The Mg to O abundance 
ratio (by number) for region A is 0.19$^{+0.10}_{-0.09}$ for the 
single plane shock and 0.44$^{+0.22}_{-0.25}$ for the two component 
(plane shock and Sedov) model. Compared to the LMC and solar abundance 
ratio of Mg/O $\sim$ 0.05, these support a significant overabundance of 
Mg in the interior of N49B.

\section{\label{sec:disc} DISCUSSION}

The overabundant Mg and Si in the center of N49B suggest the presence
of SN ejecta within the SNR's interior. It is remarkable that
metal-rich ejecta can still be significant and observable in an aged
SNR ($\sim$10$^4$ yr, Hughes et al. [1998]). N49B thus joins a growing
number of middle-aged SNRs showing evidence for ejecta (e.g., Hughes
et al. 2003). Mg is produced in astrophysical sites such as hydrostatic 
C and Ne burning, explosive C/Ne-burning during core-collapse SNe 
\citep{thielemann96}, and thermonuclear (Type Ia) SNe \citep{iwamoto99}. 
In nucleosynthesis models, though, Mg is accompanied by other elemental 
species with O invariably being produced in greater quantities. The 
presence of ejecta rich in Mg, but not in O/Ne, is thus surprising. 
It is possible that particular heating or cooling effects have produced 
a thermal/ionization state in which O/Ne line emission is suppressed 
relative to Mg, but a full investigation of such plasma conditions is 
beyond the scope of this paper. We note that N49B is not classified as 
an O-rich SNR in the optical band, nor have there been any reports of 
ejecta in this SNR prior to this work.

\begin{figure}[t]
\figurenum{2}
\centerline{\includegraphics[angle=-0,width=3.5in]{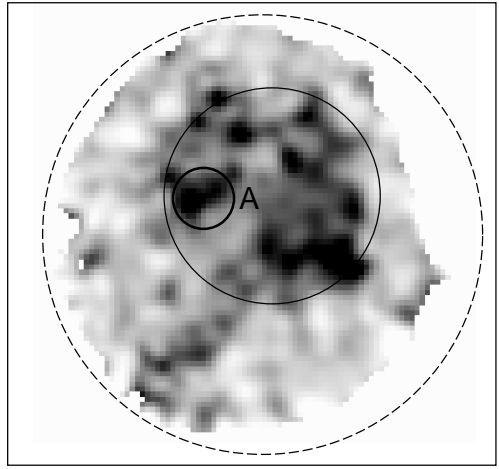}}
\figcaption[]{Gray-scale Mg EW image. Darker gray-scales means higher EWs. 
The photon energy bands used to generate this EW image are 1280 $-$ 1440 
eV for the line emission, 1140 $-$ 1240 eV and 1550 $-$ 1700 eV for the 
estimation of the underlying continuum. The dashed circle displays a fiducial 
boundary for the SNR as presented in Figure \ref{fig:fig1}. The thin solid 
circle is the 0$\farcm$7 radius region where the total Mg mass was estimated.
The Mg EW is $\sim$300 $-$ 700 eV, typically interior to the encircled area 
and $<$100 eV otherwise. The thick small circle is region A. In order to 
remove the noise due to the background, EWs are set to zero where the 
estimated continuum fluxes are low ($<$5\% of the maximum). The line and 
continuum images were extracted with 2$\arcsec$ pixels and smoothed by a 
Gaussian with $\sigma$ = 4$\arcsec$.
\label{fig:fig2}}
\end{figure}

\begin{figure}[h]
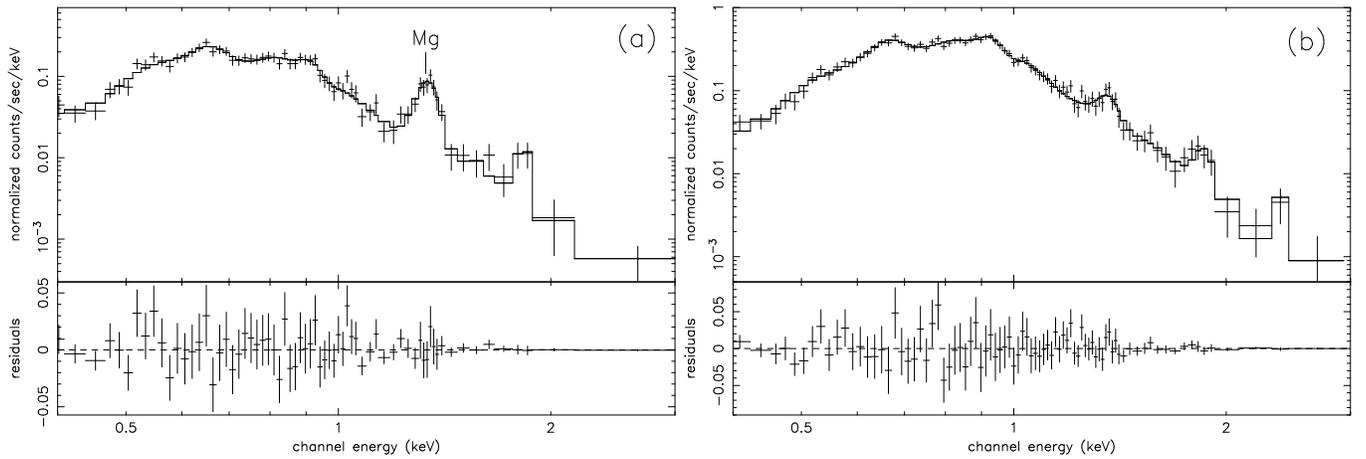

\figurenum{3}
\centerline{\includegraphics[angle=-90,width=3.5in]{fig3a.ps}
\includegraphics[angle=-90,width=3.5in]{fig3b.ps}}
\figcaption[]{(a) The spectrum from a high Mg EW region (region A).
(b) The spectrum from a bright X-ray filament (region B).
\label{fig:fig3}}
\end{figure}

We estimate the total mass of Mg in N49B by determining the density from 
spectral fits of region A and then applying it to the entire volume of 
the Mg-rich region. For region A, we assume emission from a spherical 
volume $V$ = 3 $\times$ 10$^{57}$ $f$ $d^{3}_{50}$ cm$^{3},$ where $f$ 
is the X-ray volume filling factor and $d_{50}$ is the distance to the 
SNR in units of 50 kpc. Using the  best-fit volume emission measure ($EM$) 
and Mg abundance (Table \ref{tbl:tab1}), we find $n_e$ $\sim$ 0.07 
cm$^{-3}$ under the assumption that the bulk of the electrons are from Mg 
(i.e., a {\it pure} Mg ejecta) where $n_e$ $\sim$ 10$n_{Mg}$ for the mean 
charge state of Mg implied by the measured temperature and ionization 
timescale. Assuming a spherical volume with a radius of 0$\farcm$7 (large 
solid circle in Figure \ref{fig:fig2}) for the {\it entire} enhanced Mg 
emission region, we find a total Mg ejecta mass of $\sim$18$f^{1/2}$ 
$M_{\odot}$ which is a conservative upper limit. For the case where the 
electrons are primarily from H/He, this becomes $\sim$ 0.5$f^{1/2}$ 
$M_{\odot}$. There is some uncertainty associated with extrapolating the 
density obtained from within a small area directly to the larger area of 
the metal-rich ejecta. The broadband surface brightness of the Mg-enhanced 
region is however relatively uniform, indicating a modest $EM$ variations 
over the larger area. Our simple extrapolation thus provides a reasonable, 
although crude, estimate of the plasma density. The large mass of Mg we 
derive, even with the large uncertainties, strongly suggest a 
core-collapse SN from a massive progenitor ($>$ 25 $M_{\odot}$) 
\citep{thielemann96}.

The bright filaments along the periphery of the SNR are not evident in 
the EW maps, and their spectra can be fitted with LMC abundances. These 
filaments also appear to be compressed, and have little faint X-ray 
emission beyond them (except notably for the western side). These bright 
features most likely represent interactions of the blast wave with dense
portions of the ISM. Because of the ambiguous geometry for these thin 
filaments, estimates of the electron density for these regions 
are difficult. We assume a slab-like cylindrical geometry for the emitting 
volume for region B. For the apparent angular size of 30$^{\prime\prime}$ 
$\times$ 8$^{\prime\prime}$, and a line-of-sight path-length of $\sim$3 pc 
comparable to an angular scale of $\sim$15$^{\prime\prime}$ at $d$ = 50 kpc, 
we obtain $V$ $\sim$ 1 $\times$ 10$^{57}$ cm$^{3}$ (for $f$ $\sim$ 1), 
and $n_e$ $\sim$ 15 cm$^{-3}$ for this filament. The derived X-ray 
emitting mass is $\sim$8$M_{\odot}$. Considerable mass has thus been 
encountered along the filamentary regions. 

We detect a faint shell-like feature extending beyond the bright 
X-ray filament on the western side of N49B. This feature has a fairly 
large angular size ($\sim$1$\farcm$5 $\times$ 0$\farcm$5) with 
soft emission, and is spatially coincident with the radio 
shell on the west. These  morphologies suggest that this 
feature is the blast wave shock front propagating into less dense portions 
of the ISM, whereas the bright X-ray filaments are emission from the shock 
encountering denser ISM. This may also be the case in the south-east as 
shown by the faint diffuse X-ray emission beyond the bright X-ray 
knots. These ``superpositions'' of bright and faint features appear to 
be projection effects of the blast wave expanding into a cloudy ISM. The
presence of bright optical knots corresponding to the bright X-ray
filaments in the southwest and in the middle of the eastern side 
supports this hypothesis. We can get an estimate of the
ambient density contrast by comparing the surface brightness of the
X-ray faint and bright features. Assuming similar spectra and 
line-of-sight depths for these regions, the surface brightness
ratio implies the ratio of electron density squared. We compare the
bright X-ray filament on the western side of the SNR to the faint
X-ray shell just ahead of it, where the surface brightness ratio is
$\sim$15. The bright filament has $n_e$ $\sim$ 5 cm$^{-3}$ (derived by
actual spectral fitting), which implies $n_e$ $\sim$ 0.7 cm$^{-3}$ for 
the faint X-ray shell. This latter value is consistent with previous 
estimates of the average preshock density around N49B \citep{hughes98}.
The blast wave thus appears to be expanding into an ambient ISM with local 
density variations of an order of magnitude.

N49B appears to lie in projection near an H{\small II} region
suggesting a core-collapse SN. However, we find no apparent evidence 
for the associated neutron star or its wind nebula. The radio data also 
showed no compact source or strong polarization within N49B
\citep{dickel98}. We estimate a flux limit for an embedded point source.
We convolved a point source, assuming a power law spectrum and
various fluxes, with the ACIS-S3 point spread function. Each simulated
point source was then added to the central 30$^{\prime\prime}$ $\times$ 
30$^{\prime\prime}$ observed region of N49B. A 3$\sigma$ detection limit 
is then $\sim$30 counts, corresponding to an X-ray luminosity of $L_X$ 
$\sim$ 2 $\times$ 10$^{33}$ ergs s$^{-1}$ (0.5 $-$ 5 keV). This limit 
does not conclusively rule out the presence of an embedded neutron star 
in N49B, since some Galactic spin-powered pulsars have $L_X$ $\la$ 
10$^{33}$ ergs s$^{-1}$ (e.g., Becker \& Tr\"umper 2002).

\acknowledgments
This work has been in parts supported by NASA contract NAS8-01128. 
JPH was supported by {\it Chandra} grants GO1-2052X, GO2-3068X, and 
GO2-3080B. POS was supported by NASA contract NAS8-39073.

\begin{deluxetable}{ccccccccccc}
\rotate
\footnotesize
\tablecaption{Results from the Spectral Model Fittings\tablenotemark{a}.
\label{tbl:tab1}}
\tablewidth{0pt}
\tablehead{ \colhead{Region} & \colhead{$N_H$} & \colhead{$kT$} & 
\colhead{$n_et$}& \colhead{O} & \colhead{Ne} & \colhead{Mg} & \colhead{Si} &
\colhead{Fe} & \colhead{$EM$\tablenotemark{b}} & $\chi^{2}$/$\nu$ \\
 & \colhead{(10$^{21}$ cm$^{-2}$)} & \colhead{(keV)} & 
\colhead{(10$^{11}$ cm$^{-3}$ s)} & & & & & & & }
\startdata
A & 2.3$^{+1.1}_{-2.0}$ & 0.36$^{+0.14}_{-0.09}$ & 4.5$^{+4.1}_{-2.4}$ & 
0.41$^{+0.41}_{-0.15}$ & 0.37$^{+0.47}_{-0.17}$ & 1.73$^{+1.92}_{-0.68}$ & 
1.44$^{+1.63}_{-0.71}$ & 0.20$^{+0.23}_{-0.08}$ & 2.9$^{+1.5}_{-1.5}$ & 
43.3/58 \\
B & 3.3$^{+1.6}_{-1.3}$ & 0.34$^{+0.14}_{-0.10}$ & 11.2$^{+4.7}_{-3.4}$ & 
0.33$^{+0.10}_{-0.07}$ & 0.34$^{+0.12}_{-0.09}$ & 0.28$^{+0.13}_{-0.10}$ & 
0.42$^{+0.27}_{-0.21}$ & 0.15$^{+0.05}_{-0.03}$ & 14.7$^{+2.9}_{-2.9}$ & 
47.1/75 \\
\enddata
\tablenotetext{a}{Errors are with a 90\% confidence. Errors on abundances
are obtained after fixing $N_H$, $kT$, and $n_et$ at the best-fit values.}
\tablenotetext{b}{Volume emission measure in units of 10$^{58}$ cm$^{-3}$, 
$EM$ = ${\int}n_en_HdV$.}
\end{deluxetable}

\end{document}